\documentstyle[amstex,epsfig,graphicx,subfigure,titlepage,12pt]{article}

\begin{document}

\renewcommand{\thesection}{\Roman{section}.} \baselineskip=24pt plus 1pt
minus 1pt

\begin{titlepage}
\vspace*{0.5cm}
\begin{center}

\LARGE\bf Talbot Effect with Matter Waves
\\[1.5cm]
\normalsize\bf Farhan Saif
\end{center}


\vspace{7pt}
\begin{description}


\item  []Department of Electronics, Quaid-i-Azam University, Islamabad 45320, Pakistan.
\item  []Center for Applied Mathematics and Physics, National University of Science and Technology, Islamabad, Pakistan.



\end{description}

\vspace{0.3cm}

\normalsize {Talbot effect in the \textit{space-time} evolution of matter waves is analyzed and shown 
that the matter waves at relativistic and non-relativistic velocities exhibit coherence beyond the grating and display Talbot self-imaging.  
The grating is realized by considering equidistant narrowly peaked coherent probability distributions at the onset.
Theoretical frame work is based on a quantum particle at relativistic and non-relativistic velocities in a one dimensional box.
The matter waves, in their evolution, display Talbot self-imaging and fractional self-imaging as restructuring, respectively, at
Talbot length and fractional Talbot lengths.  }

\hspace{0.4cm}









\end{titlepage}

\newpage


\section{Introduction}

The Talbot Effect in classical optics, observed in 1836 by Henry Fox Talbot,
is a near-field diffraction effect~\cite{talbot}. As a laterally 
periodic coherent wave is incident on a diffraction grating, the grating 
image repeats at regular distances away from the grating plane. The regular 
distance is called the Talbot Length. At fractions of the Talbot length we find 
fractions of the grating images~\cite{Rayleigh}. 
The phenomenon has applications in optical metrology, photolithography, 
optical computers~\cite{Patorski}, as well as, in data compression~\cite{jebali}
electron optics and microscopy~\cite{Cowley}. Sir Michael Berry and Susanne Klein 
have shown a relation of Talbot effect with quantum revival phenomenon of a 
particle~\cite{berry}. 
Later, in another seminal paper a close relavance between 
Talbot effect with light and quantum revivals of atoms 
in a one dimensional box was developed~\cite{berry2001}.
Integer and fractional nonlinear Talbot 
effect is recently reported in nonlinear photonic crystals~\cite{zhang}. In atom optics
the effect is seen using atoms with non-relativistic 
velocities, displaying self imaging at Talbot lengths~\cite{chapman}. 
Particle in a box has been treated
as playground to study quantum revivals~\cite{robinett,saif} and space-time dynamics which weaves quantum 
carpets as a result of delicate 
interference between eigen modes of a system~\cite{Kaplan2000,Marzoli1998}. 
For a slightly relativistic particle in a one dimensional box, quantum carpets are explained earlier by using
Green's function where slight relativistic corrections appear in times of
revivals~\cite{Marzoli2008}. One dimensional Dirac equation is
used to treat a relativistic particle along a circle~\cite{Strange2010}. In this contribution,
I explain Talbot effect for matter waves at non-relativistic and relativistic velocities. 
Theoretical treatment is conceptually based on the evolution of a quantum particle in a one dimensional box,
which, serves as a finite universe model to explain Talbot effect. 
I show that the matter waves manifest interference leading to restructuring 
at times which determine self-imaging of Talbot. Furthermore, I illustrate that matter waves with relativistic velocities 
display unique Talbot carpets in space-time evolution, different from classical optics.
Hence, I provide a rigorous treatment to the problem based on relativistic quantum mechanics, first,
to understand Talbot effect with matter waves and, second,
to explain the evolution of a relativistic particle in a one dimensional box. 

\begin{figure}[tbp]
\centerline{\includegraphics[scale=.55]{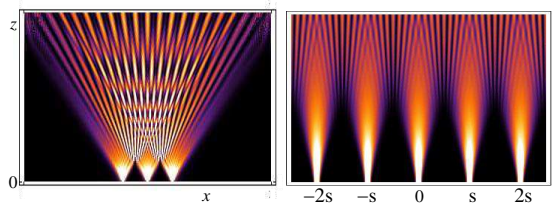}}
\caption{
{\it Left panel} shows that matter waves passing a few-slit grating with non-relativistic velocity, $v$, display interference pattern but no Talbot self-imaging. {\it Right panel} describes that matter
wave passing a diffraction grating, defined mathematically by equi-distant narrowly-peaked probability distributions, 
display Talbot fringes even in early evolution. In the early evolution individual distributions 
after respective initial quantum dispersion in free space interfere with each other. 
Here, grating spacing $s=L/10$, grating elements widths $\Delta=0.001L$, $z=vt$, where $L$ is the length of the grating and $t$ is evolution time.}
\label{Fig3}
\end{figure}
In order to explain Talbot self imaging for matter waves, let us consider an initial array of equidistant narrowly peaked probability 
densities in a 1D box, which serves as equidistant wave-fronts emerging from grating. 
In order to observe Talbot effect, the distance between two wave-fronts, which is suitably regarded as grating-distance, 
is taken much smaller than the length of the box. During its propagation the array displays 
quantum interference, for a very early evolution, see Fig.~\ref{Fig3}(right panel).  
The probability density for the matter wave at position, $x$, and time, $t$, is calculated as 
\begin{eqnarray}
W(x,t) = \int_{0}^{L}dx^{\prime }\int_{0}^{L}dx^{\prime \prime }\,K(x,t|x^{\prime
},x^{\prime \prime}) \omega^\dagger(x^{\prime })\,\omega(x^{\prime \prime })
\label{eq4}
\end{eqnarray}
and, therefore, follows from the integration of the kernel 
\begin{equation}
K(x,t|x^{\prime },x^{\prime \prime})=G^\dagger(x,t|x^{\prime })\,G(x,t|x^{\prime\prime }).  
\label{eq5}
\end{equation}
Here, $\omega=\omega(x,t=0)$ denotes the initial wave function and, $G$ expresses
Green's function, defined as,
\begin{equation}
G(x,t|x^{\prime })\equiv \sum_{m=1}^{\infty }\varphi_{m}(x)\,\varphi_{m}(x^{\prime
})e^{-iE_{m}t/\hbar },  
\label{eq3}
\end{equation}
written in terms of the energy eigen functions $\varphi_{m}$ and energy eigen values, $E_{m}$. 
The evolution of a particle of mass $M$ in one dimensional box of length $L$, along $x-$axis, may be expressed by 
the one-dimensional relativistic Hamiltonian,  
$H=\sqrt{(Mc^{2})^2+({\hat p}_x c)^2}-Mc^2$.  
In the limit $c\rightarrow \infty $ the Hamiltonian $H$ reduces to 
non-relativistic form, whereas, retaining relativistic corrections to order $1/c^2$ provides slightly relativistic Hamiltonian. 
The solution, $\varphi(p,x)=\mathcal{N} \sin(px/\hbar)$, satisfies the equation
$\left\{ {{\hat H} - E_p } \right\} \varphi(p,x) = 0$, in non-relativistic and 
slightly relativistic limits~\cite{Marzoli2008},
where $p=\hbar k=m\pi\hbar/L$, 
and $\mathcal{N}$ is normalization constant.

\section{Talbot effect with relativistic matter waves}

In relativistic case we solve the Dirac equation for one dimensional box, that is, 
\begin{equation}
H\varphi=(\alpha_z{\hat p}_x c +\beta Mc^2))\varphi=E\varphi,
\end{equation}
where $\varphi$ denotes two-component spinor depending upon $x$, belongs to $[0,L]$~\cite{alberto}, and $\alpha_z$ and $\beta$ are the matrices
\begin{equation}
\alpha_z= \left(
\begin{array}{cc}
0 \,\,\,
\sigma_z \\
\sigma_z \,\,\,
0
\end{array}
\right)\;, \notag
\hspace{1cm}
\beta= \left(
\begin{array}{cc}
I \,\,\,
0 \\
0 \,\,\,
-I
\end{array}
\right).
\end{equation}
Here, $\sigma_z$ is Pauli matrix and $I$ is unit matrix. The solution of the Dirac equation
within the box is  
\begin{equation}
\varphi_m=\mathcal{N} e^{i\delta/2} \left(
\begin{array}{cc}
2\cos\left(\frac{m\pi x}{L}-\frac{\delta}{2}\right)\chi\\
2iP\sin\left(\frac{m\pi x}{L}-\frac{\delta}{2}\right)\sigma_z\chi
\end{array}
\right)
\label{wfunc}
\end{equation}
where, the constant $\mathcal{N}$ is obtained from the normalization condition. In addition, we write, 
\begin{eqnarray}
\delta_m=\arctan\left(\frac{2P_m}{1-P_m^2}\right)\,\,\,\,\,\,\, P_m=\frac{\hbar kc}{E+Mc^2}.
\end{eqnarray} 
Moreover, 
$\chi$ is an arbitrary two-component normalized spinor, i.e. $\chi^\dagger\chi=1$, 
and  
\begin{equation}
E_{m}=Mc^2\sqrt {1+(2mq)^2}.
\label{eq14}
\end{equation}
The controlling parameter $q$ is defined as,  
$q= \left( \frac{2\pi \hbar }{Mc}\right) /4L= \lambda_c/4L =v/c $, a ratio between 
Compton wavelength $\lambda_c$ and four times the length of the box, or a ratio between effective velocity $v$ and speed of light, $c$. 
In order to understand the space-time dynamics, we write the Green's function for the component which follows sinusoidal dependence on position, $x$,  
that leads to the calculation of the kernel, viz.,
\begin{eqnarray}
K = D_r( +,+) + D_r(-,-)- D_r(+,-) - D_r(-,+),
\label{kdsr}
\end{eqnarray}
Here, we use scaled coordinates $\xi=x/L$ and $\tau=t/T$, in addition ${\cal E}_m=TE_m/\pi\hbar$ and $T=4ML^2/\pi\hbar$, and
\begin{eqnarray}
&&D_r(+,+)=D_r\left( {+\eta ,+\zeta } \right) = \sum\limits_{m,n = - \infty }^\infty
e^{i\pi \left( {m\eta - n\zeta } \right)} f_{+,+}(\delta) \notag \\
&& \cdot e^{  i\pi \left\{ \left( {m - n} \right)\xi - (TMc^2/\pi\hbar)\{\sqrt{1+(2mq)^2}-\sqrt{1+(2nq)^2}\}\tau
\right\} }.
\end{eqnarray}
Here, $f_{+,+}=1$, however, $f_{-,-}=e^{i({\rm sgn}(m)\delta_m-{\rm sgn}(n)\delta_n)}$, $f_{+,-}=e^{-i {\rm sgn}(n)\delta_n}$, and $f_{-,+}=e^{i {\rm sgn}(m)\delta_m}$. In exponent ${\rm sgn}$ defines signum function.

The double series transform 
\begin{eqnarray}
\sum\limits_{m,n = - \infty }^\infty f_{m,n} = \frac{1}{2}\sum\limits_{j,l
= - \infty }^\infty ( - 1)^{jl}  
\int\limits_{ - \infty }^\infty d\rho \cdot f\left[ \frac{1}{2}\left( j +
\rho \right),\frac{1}{2}\left( j - \rho \right) \right]e^{i\pi l\rho }
\end{eqnarray}
provides the following expression for $D_r \left( {\eta ,\zeta }
\right)$, {\it i.e.}, 
\begin{eqnarray}
D_r(+,+) = \sum\limits_{l = - \infty }^\infty \sum_{j=1}^{\infty}(-1)^{jl}  \left[
 e^{i\pi j\frac{\eta - \zeta }{2}} {\cal I}_+   +
 e^{-i\pi j\frac{\eta - \zeta }{2}} {\cal I}_-\right] 
+ \sum\limits_{l = - \infty }^\infty {\cal I}_{0l},
\label{gdsr}
\end{eqnarray}
where 
\begin{eqnarray}
{\cal I}_{\pm}&=&\int_0^{\infty} d\rho \cos\left(\pi\psi\rho \pm \alpha \{ s_+(\rho) - s_-(\rho) \} \right),\notag\\
{\cal I}_{0l}&=& \int_0^\infty d\rho \cos(\pi\psi\rho).\notag
\end{eqnarray}
In order to simplify the expressions here we use the following abbreviations: $\psi= \frac{\eta+\zeta}{2}+\xi +l$;  
$\alpha = Mc^2T/\pi\hbar$; and $s_{\pm}(\rho)=\sqrt{1+q^2(j \pm \rho)^2}$.

I solve the integrals, ${\cal I}_{\pm}$, and substitute the expressions in Eq.~(\ref{eq4}). Hence, for a matter wave with finite initial distribution, $\Delta$, much smaller than the grating spacing, $s$, the probability density becomes, 
\begin{eqnarray}
&&W(\xi,\tau)=\mathcal{N}^4\frac{1}{2q}\sum_{l=-\infty}^{\infty} \sum_{j=1}^{\infty} (-1)^{jl} e^{-j^2\pi^2\Delta^2} \notag\\
&& \,\,\,\,\, \times \sum_{n,m=-\infty}^{\infty}  \sum_{k_1,k_2=-k}^k i^{n+m} \,I_{nm}^{(z_0)}\notag\\
&&\,\,\,\,\left[{\cal J}_-(k_1,k_2)\left(e^{i\Omega_+}\Gamma_+^{(+)}+(-1)^{n+m}e^{-i\Omega_+}\Gamma_-^{(+)}\right)\right. \notag\\
&&\,\left. +{\cal J}_+(k_1,k_2)\left(e^{i\Omega_-}\Gamma_+^{(-)}+(-1)^{n+m}e^{-i\Omega_-}\Gamma_-^{(-)}\right)
\right].
\label{wxtf}
\end{eqnarray}
Here, I introduced ${\cal J}_-(k_1,k_2)=J_m(\gamma_-)e^{i\pi j\frac{k_1 - k_2}{2}}$ and ${\cal J}_+(k_1,k_2)=J_m(\gamma_+)e^{-i\pi j\frac{k_1 - k_2}{2}}$.
Furthermore, $k_1$ and $k_2$ are integers and $k$ corresponds to total number of initial narrowly peaked probability densities at grating spacing, 
$s$~\cite{onedb}. Furthermore, $\Omega_\pm= j\pi(\xi\pm 2\tau/q+l\pm (k_1+k_2)/2)$, and $\Gamma_\pm^{(\pm)}$ is defined as
\begin{eqnarray}
\Gamma_\pm^{(\pm)}=\frac{q}{\pi}\int_{-\infty}^{\infty}d{\chi} e^{-\frac{q^2}{2\Delta^2\pi^2}({\chi}- a_0^\pm)^2}I_n({\chi} + \chi_0^{(\pm)}).
\label{gam}
\end{eqnarray}
The parameter $a_0^\pm=\pi(k_1+k_2+ 2ij\pi\Delta^2)/2q$ and $\chi_0^{(\pm)}= \pi(\xi \pm \tau/q+l+\frac{k_1+k_2}{2})/q$. The four terms in 
the expression for $K$ in Eq.~(\ref{kdsr}) pick four different corresponding combination of signs for $k_1$ and $k_2$. Hence, the full expression becomes 
four-fold to what we have in Eq.~(\ref{wxtf}).

The final expression though difficult in its structure is easy to analyze. 
As discussed above $n$ and $m$ satisfy the equation $n+m+2\ge 0$, keeping in view minimum criteria we choose $n=-2$ and $m=0$. Furthermore, we neglect  imaginary part in $a_0^{\pm}$ as $\Delta^2<<1$, that simplifies the expression for $W(\xi,\tau)$, as it becomes
\begin{eqnarray}
&&W(\xi,\tau)=\mathcal{N}^4\frac{1}{2q}\sum_{l=-\infty}^{\infty} \sum_{j=1}^{\infty} (-1)^{jl+1} e^{-j^2\pi^2\Delta^2}I_{-2,0}^{(z_0)}\notag\\
&& \,\,\, \times 
\sum_{k_1=-k}^k  
\left[(J_0(\gamma_+)\cos\Omega_-\Gamma^{(-)}  + J_0(\gamma_-)\cos\Omega_+\Gamma^{(+)}\right]. 
\label{swxtf}
\end{eqnarray}
For simplicity I consider $k_1=k_2$.
Here $J_0(\gamma_\pm)$ is a function of time, and $\cos\Omega_\pm$ defines the space-time structures. Furthermore, solving Eq.~(\ref{gam}) in the vicinity of $\chi_0^{(\pm)}$, we get,
\begin{eqnarray}
\frac{\Gamma^{(\pm)}}{\sqrt{2\pi}\Delta}=I_n(\chi_0^{(\pm)})-\frac{\chi_0^{(\pm)}}{2}(I_{n-1}(\chi_0^{(\pm)})+I_{n+1}(\chi_0^{(\pm)})),
\end{eqnarray}
which is independent of $j$. 
 
With the help of asymptotic expression $J_{0}(z)\approx \sqrt{2/\pi z}\cos z$, we get 
\begin{eqnarray}
J_0(\gamma_{\pm})\cos\Omega_\mp=\cos(j\pi(\xi\mp (2/q\mp 1/(j\pi q^2) )\tau+l)).
\end{eqnarray}
Note that for all the values of $j$ we have periodic behavior both in space and in time. In $(\xi, \tau)$ plane there exists maximum probability lines 
with slope $\mp (2/q\mp 1/(j\pi q^2) )$. It is interesting to note that the $j$ dependent term in the expression for slope becomes negligibly small for $q\ge 1$, however, for smaller $q$ values it contributes, removes degeneracy, and support many canals and ridges behavior corresponding to different $j$ values. Hence, for $q=1$ and higher, we effectively find higher probabilities following straight lines with slope $\pm q/2$ for all values of $j$, which manifests quantum coherence at relativistic scale.
The analytical results show a good comparison with exact numerical results.
For the component of Dirac solution, given in Eq.~(\ref{wfunc}), which does not satisfy the boundary conditions we get similar results for probability density with a difference of phases between constituent terms corresponding to $D_r$'s, given in Eq.~(\ref{kdsr}).

\section{Talbot effect with non-relativistic matter waves}

As the particle evolves with non-relativistic velocity, the corresponding expression for $D_r(+,+)$ becomes
\begin{eqnarray}
\sum_{j,l=-\infty}^{\infty} (-1)^{jl} e^{i\pi j \frac{(\xi'-\xi'')}{2} }\delta\left(l+\xi-2j\tau+ \frac{\xi'+\xi''}{2}\right).
\end{eqnarray}
The probability density, as calculated following Eq.~(\ref{eq4}), reveals
\begin{eqnarray}
W(\xi,\tau)&=&\mathcal{N}^4\sum_{l=-\infty}^{\infty}\sum_{j=\pm 1}^{\pm\infty} (-1)^{jl} e^{-\frac{j^2\pi^2 \Delta^2}{2}}\notag\\
&\times & \sum_{k_1,k_2=-k}^k e^{i\pi j\frac{k_1 - k_2}{2}} e^{-\frac{1}{2\Delta^2} \left(\xi - 2j\tau +l + \frac{k_1+k_2}{2}\right)^2}\notag\\
&+& \sum_{l=-\infty}^{\infty}\sum_{k_1,k_2=-k}^k  e^{-\frac{1}{2\Delta^2} \left(\xi +l + \frac{k_1+k_2}{2}\right)^2}.
\label{nrdist}
\end{eqnarray}
As in relativistic case, the four terms in Eq.~(\ref{kdsr}), respectively, take one corresponding sign combination for $k_1$ and $k_2$.
Hence, the complete expression for $W$ is composed of four similar terms, given in Eq.~(\ref{nrdist}), with four different combination of signs for $k_1$ and $k_2$. Beyond $\tau=0$, a jet of sub-distributions emerges following straight lines of slope $2j$ in ($\xi,\tau$) plane and shifted in time by $l\pm\frac{k_1\pm k_2}{2}$. We note that the dominant values of $j$ which contribute in defining the probability distribution as a function of space and time have width, $1/\sqrt{2}\pi\Delta$. For the reason smaller value of $\Delta$ corresponds to more straight lines which define richer space time structures. Hence we find, for each straight line with slope $2j$ there exists a counterpart with slope $-2j$ and their interference carves carpet structures, with a periodicity in time. 
The last summation term in Eq.~(\ref{nrdist}), explicitly written for $j=0$, corresponds to straight lines with zero slope, along the $\tau$ axis, for $l$ and $\pm \frac{k_1\pm k_2}{2}$.

\begin{figure}[tbp]
\centerline{\includegraphics[scale=.55]{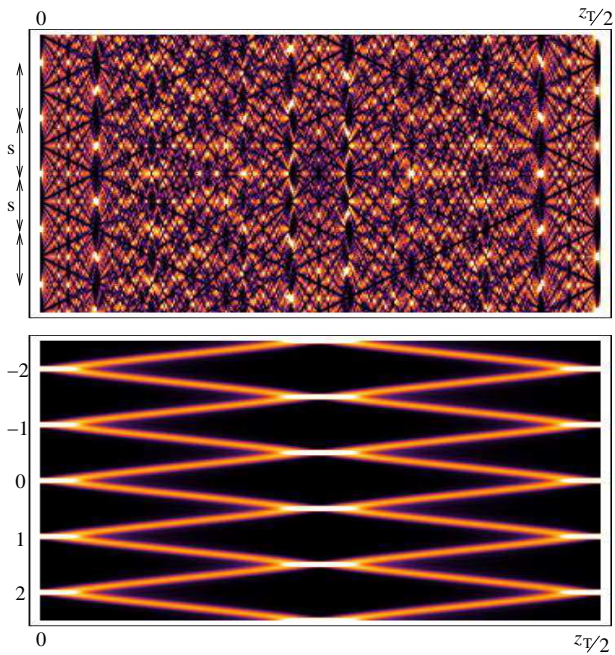}}
\caption{Talbot self-imaging for matter waves with non-relativistic velocity (upper panel). At fraction of Talbot length we note fraction 
of grating images. For matter waves with relativistic velocity (lower panel) we find simpler Talbot carpet. Since for each initial wave-let there appears two symmetric non-zero probability density lines, explained in the text, the final Talbot carpet for matter wave is different from non-relativistic one and, that for {\it e. m.} waves in classical optics. The Talbot length, however, is larger now as compared with non-relativistic case due to time dilation effect. The parametric values are the same as in Fig.~\ref{Fig3}.
}
\label{Fig4}
\end{figure}

Matter waves with non-relativistic velocity display Talbot self-imaging, similar to light waves in classical optics, shown in Fig.~\ref{Fig4} (upper panel). 
The space time dynamics for non-relativistic matter waves is defined by Eq.~(\ref{nrdist}), which depicts that the quantum Talbot carpets are more dense in structure as compared with quantum carpets for a particle in a box, as there exist $4k(k+1)$ 
more low and high probability density lines in quantum Talbot carpets, respectively,
named as canals and ridges. Here, $k$ is the total number of initial probability densities or wave-fronts which define grating. 
The self-imaging takes place at Talbot length, $z_T=2\lambda_{dB}^{-1}s^2$~\cite{Rayleigh}, and fractional imaging occurs at fraction of the distance. Here, $s$ is the grating's spacing and $\lambda_{dB}$ is the 
de-Broglie wavelength of the matter waves, defined as $\lambda_{dB}=2\pi\hbar/Mv=(T\frac{v}{8L^2})^{-1}$, as the particle moves with effective velocity $v$, and $T$ is the time of quantum revival. 

Matter waves with relativistic velocities display self imaging in complete contrast to non-relativistic case, and to {\it e.m.} waves in classical optics. 
As discussed above, Eq.~(\ref{swxtf}), in relativistic case there are only two possible non-zero probability density lines in ($\xi,\tau$) plane, and wave packet intersect each other
periodically, thus, defining full restructuring of original wavelets. This leads to self-imaging of the grating at Talbot lengths.  Whereas at any intermediate time there occurs self-imaging twice of the grating structure, see Fig.~\ref{Fig4}(lower panel). 
We defines the Talbot length, $z_T=2\lambda_{dB}^{-1}s^2=2(T^{rel}\frac{v}{8s})^{-1}$. 
Hence, in relativistic case Talbot carpet is simpler in structure, as compared with classical optics, however Talbot length is longer due to time dilation relativistic correction~\cite{trel}.

\section{Conclusions}

I conclude that in the present contribution, I develop the theory of Talbot effect for matter waves with relativistic and non-relativistic velocities.
I show that quantum coherence prevails and drastically modifies the interference behavior in relativistic case which results in an exact Talbot imaging, available in non-relativistic quantum mechanics as well, however, following different Talbot carpets. The mathematical analysis provides theoretical understanding of a relativistic particle in one dimensional box as well. 
Keeping in view latest technological development, the analytical results 
can be realized using cold neutral atoms~\cite{chapman}, and in recent experiments with electrons~\cite{cronin}.

The author acknowledges partial funding from Pr\'o-Reitoria de Pesquisa-UNESP, Sao Paulo, and HEC NRP 20-1374.



\begin{thebibliography}{99}
\bibitem{onedb} For a particle in a box problem the summation over $k_1$ and $k_2$ disappears 
and we consider $k_1=k_2=k_0$, where $k_0$ ranges from 0 to 1 and describes particle position within box.
\bibitem{trel} As the particle follows relativistic velocity the time of its revival becomes
\begin{equation}
T^{rel}=T\left(\frac{(1+(2m_0q)^2)^{3/2}}{1-(2m_0q)^2}\right).
\end{equation}
\end{thebibliography}
\end{document}